\documentclass{nature}

\usepackage{graphicx}
\usepackage{multirow}
\graphicspath{ {images/} }
\DeclareGraphicsExtensions{.png,.pdf,.eps}

\usepackage{hhline}
\usepackage{epstopdf}
\usepackage{amsmath,amssymb}

\usepackage[colorlinks,urlcolor=cyan,citecolor=blue,linkcolor=blue]{hyperref}

\newcommand{\aj}{Astron. J.}

\newcommand{\araa}{Annual Review of Astronomy and Astrophysics}
\newcommand{\apj}{Astrophys. J.}
\newcommand{\apjl}{Astrophys. J., Letters}

\newcommand{\aap}{Astron. Astrophys.}

\newcommand{\icarus}{Icarus}

\newcommand{\mnras}{Mon. Not. R. Astron. Soc.}

\newcommand{\pasp}{Publ. Astron. Soc. Pacific}

\newcommand{\psj}{The Planetary Science Journal}

\newcommand{\ssr}{Space Science Reviews}

\newcommand{\nat}{Nature}

\newcommand{\gca}{Geochimica Cosmochimica Acta}
\newcommand{\grl}{Geophysics Research Letters}

\newcommand{\jgr}{Journal of Geophysics Research}

\usepackage{xcolor}

\title{Atmospheric carbon depletion as a tracer of water oceans and biomass on temperate terrestrial exoplanets}

\author{
Amaury H.M.J. Triaud$^{1,+,}$\thanks{corresponding authors ({\color{blue}a.triaud@bham.ac.uk} and {\color{blue}jdewit@mit.edu})}, 
Julien de Wit$^{2,+,*}$, 
Frieder Klein$^{3}$, 
Martin Turbet$^{4}$, 
Benjamin V.\ Rackham$^{2,5}$,  
Prajwal Niraula$^{2}$, 
Ana Glidden$^{2,5}$, 
Oliver E. Jagoutz$^{2}$, 
Matej Pe\v c$^{2}$, 
Janusz J. Petkowski$^{2,6,7}$, 
Sara Seager$^{2,8,9}$, 
Franck Selsis$^{10}$ 
} 

\begin{document}
\maketitle

\textsl{$^+$These authors contributed equally to this work.}

\begin{affiliations}
\item School of Physics and Astronomy, University of Birmingham, Edgbaston, Birmingham B15 2TT, United Kingdom;
\item Department of Earth, Atmospheric and Planetary Sciences, Massachusetts Institute of Technology, 77 Massachusetts Avenue, Cambridge, Massachusetts 02139, USA;

\item Department of Marine Chemistry and Geochemistry, Woods Hole Oceanographic Institution, Woods Hole, MA 02543, USA;
\item Laboratoire de Météorologie Dynamique/IPSL, CNRS, Sorbonne Université, École Normale Supérieure, PSL Research University, École Polytechnique, 75005 Paris, France;
\item Kavli Institute for Astrophysics and Space Research, Massachusetts Institute of Technology, Cambridge, MA 02139, USA;
\item JJ Scientific, 02-792 Warsaw, Poland;
\item Faculty of Environmental Engineering, Wroclaw University of Science and Technology, 50-370 Wroclaw, Poland;
\item Department of Physics, Massachusetts Institute of Technology, 77 Massachusetts Avenue, Cambridge, MA 02139, USA;
\item Department of Aeronautics and Astronautics, Massachusetts Institute of Technology, 77 Massachusetts Avenue, Cambridge, MA 02139, USA
\item Laboratoire d’astrophysique de Bordeaux, Univ. Bordeaux, CNRS, B18N, allée Geoffroy Saint- Hilaire, 33615 Pessac, France.
\end{affiliations}

\begin{abstract}

The conventional observables to identify a habitable or inhabited environment in exoplanets, such as an ocean glint or abundant atmospheric O$_2$, will be challenging to detect with present or upcoming observatories. Here we suggest a new signature. A low carbon abundance in the atmosphere of a temperate rocky planet, relative to other planets of the same system, traces the presence of substantial amount of liquid water, plate tectonic and/or biomass. We show that {\it JWST} can already perform such a search in some selected systems like TRAPPIST-1 via the CO$_2$ band at $4.3\,\rm \mu m$, which falls in a spectral sweet spot where the overall noise budget and the effect of cloud/hazes are optimal. We propose a 3-step strategy for transiting exoplanets: 1) detection of an atmosphere around temperate terrestrial planets in $\sim 10$ transits for the most favorable systems, (2) assessment of atmospheric carbon depletion in $\sim 40$ transits, (3) measurements of O$_3$ abundance to disentangle between a water- vs biomass-supported carbon depletion in $\sim100$ transits. The concept of carbon depletion as a signature for habitability is also applicable for next-generation direct imaging telescopes. 

\end{abstract}

\flushbottom
\maketitle


\section{The challenge of detecting exohabitats this decade}


In 2022, humanity entered a new era of space exploration when the very first data products from the James Webb Space Telescope (JWST) were revealed \cite{JWSTTECERS2023,Ahrer2023,Alderson2023,Feinstein2023,Rustamkulov2023}. JWST is the first of a new generation of great observatories to come online this decade and scientific accomplishments such as probing the atmospheres of terrestrial exoplanets and 3D-mapping of giant exoplanets are now within reach. 

The greatest potential for exoplanetary science lies in the detailed atmospheric characterization of a diverse range of planets to enable comparative studies informing formation, evolution, and atmospheric processes, as well as investigating and contextualizing  habitats \cite{Krissansen2022}. Thus, our community must find and then reliably study planets suited for such quests. Over the last two decades, space- and ground-based facilities have successfully delivered on this first task, yielding over 5,200 exoplanets and close to 10,000 planetary candidates (see the NASA Exoplanet Archive). With the Extremely Large Telescopes (ELTs) and JWST coming online this decade, the field of exoplanetary science is transitioning from an era of detection into one of characterization.

Observational methods developed and refined over the past two decades now allow astronomers to collect a vast host of information about an exoplanet’s atmosphere without having to leave the Solar System \cite{Deming2017}. When a planet transits in front of its host star, it is possible to derive constraints on the vertical temperature, pressure, and compositional profile of atmospheric limbs \cite{Seager2000,Brown2001,Hubbard2001,dewit2013,Madhusudhan2019}. This provides information about the materials used during planetary assembly and their subsequent orbital evolution \cite{Oberg2011,Pinhas2019}, while also revealing the current thermodynamic state of the planet \cite{Wyttenbach2020}. High-resolution spectra can distinguish molecules with different isotopes \cite{molliere2019b,Zhang2021}, measure wind velocities \cite{Snellen2010,Louden2015}, and probe evaporating exospheres \cite{Wyttenbach2017,Spake2018}. There are also methods to create crude topological and atmospheric maps for exoplanets \cite{Knutson2007,Cowan2009,Majeau2012,dewit2012,Challener2022}.

Widely applied to hot close-in gas giants \cite{Sing2016,Mansfield2022}, these methods are now increasingly in development and will soon be applied to terrestrial planets \cite{dewit2016,dewit2018,Wakeford2019,Kreidberg2019,Garcia2022}, initiating the exploration of how diverse the atmospheric and climatic properties of exoplanets are \cite{Dransfield2020} and, importantly, how those compare to the properties of planets within our Solar System. Several teams have modeled terrestrial planet atmospheres \cite{Fauchez2021,Felton2022,Turbet2022,Sergeev2022}, while others have assessed their observability \cite{Morley2017,Malik2019,Lustig2019,Wunderlich2019,Fauchez2020,Pidhorodetska2020,Fauchez2022}. Thanks to JWST and the ELTs, the time has come to move from theoretical considerations to empirical verification and exploration.

These studies find the atmospheres of transiting temperate (defined as receiving a flux between $4\times$ and $0.25\times$ that of the Earth) terrestrial planets orbiting nearby ultra-cool dwarfs like TRAPPIST-1 \cite{Gillon2017b} can be detected with JWST and their dominant atmospheric absorbers (e.g., CO$_2$ and H$_2$O) revealed with after collecting of the order of ten transits \cite{Lustig2019,Fauchez2019}. In addition, they show that the conventional biosignature gas O$_2$ \cite{Meadows2018,Catling2018} remains out of reach with JWST due to the narrowness of the O$_2$ lines combined with a low signal-to-noise ratio below $1~\rm \mu m$ even for ideal terrestrial planet targets (i.e., those orbiting late M dwarfs, because of the much reduced size of their host star and the shortness of their temperate orbits \cite{Nutzman2008, Triaud2021}). ``Biosignatures'' are defined as observables associated with the presence of life.  Even the anti-biosignature O$_2$-
X collisionally induced absorption (CIA) band at $6.4~\rm \mu m$, which is more readily accessible, remains mostly out of reach \cite{Fauchez2020}. Indeed, detecting O$_2$ would require on the order of one thousand planetary transits for the most favorable targets, which corresponds to the allowable observation time over an entire JWST cycle. It would take a few decades to collect data for planets that have orbital periods of 5 to 10 days, such as those within the habitable zone of TRAPPIST-1. In other words, with this new generation of great observatories and programs gathering up to a few hundred planetary transits of terrestrial worlds, it will be possible to reveal atmospheric constituents that are strong absorbers and later derive preliminary constraints on abundances and temperature-pressure profiles, but these observations are not expected to assess the presence of life via O$_2$ detection. The detection of O$_2$ using the ELTs appears much more promising by co-adding the individually-resolved lines of its visible and near-infrared bands (e.g., the 760-nm A-band) which takes advantage of high-resolution spectrographs \cite{Snellen2013,Rodler2014,Serindag2019}. The detection of O$_2$ \cite{Leung2020} will be attempted using transit spectroscopy (which might be difficult, \cite{Webb2023}), but also with reflected light spectroscopy (using the High-Resolution High-Contrast technique \cite{lopezMorales2019,Marconi2022}) which is available for a couple of nearby non-transiting exoplanets (e.g., Proxima b) \cite{Lovis2017,Marconi2022}. Another biosignature of interest is $\rm O_3$, produced when a large reservoir of $\rm O_2$ is UV irradiated \cite{Meadows2018,Catling2018}. $\rm O_3$ bands in the UV and optical are usually sought after for Earth-twins orbiting Solar twins, but for late-M-dwarf hosts, faint in the optical and UV, $\rm O_3$'s $4.8~\rm \mu m$ band is the most prominent feature, and of interest to this paper.


Similarly, the assessment of a planet's habitability (defined as a planet's capacity to retain large reservoirs of surface liquid water \cite{Kasting1993}) is difficult. A planet's habitability can be affected by a large range of factors, including its orbital properties, its bulk properties, the presence or lack of plate tectonics and magnetic fields, its atmospheric properties, and finally by stellar activity \cite{Kopparapu2020}presenting a complex, and thus far a mostly theoretical assessment. Empirical measurements affected by the presence of surface liquid water would be handy; however,  very few such observables of habitability (``habsignatures'', defined in a similar way to ``biosignature'' but applied to signatures of surface liquid water) exist. Examples include detecting an ocean glint \cite{Sagan1993}, as has been done for Titan \cite{Stephan2010}, but which is expected to be extremely challenging for exoplanets \cite{Lustig2018}. The presence of surface liquid water oceans could also produce a significant hydrogen exosphere, detectable in $\rm Ly\alpha$ \cite{Kameda2017,Santos2019}. However, this is a challenging habsignature too as it would not directly point to water as the related hydrogen-bearing molecule, nor whether water is present in its liquid phase \cite{Elkins2008} and is a method limited to the very nearest planetary systems due to interstellar absorption. The closest yet to a habsignature is the measure of atmospheric SO$_2$, a molecule that quickly dissolves in liquid water. Its presence in large quantities in an exoplanet's atmosphere would indicate the absence of large bodies of liquid water \cite{Loftus2019,Jordan2021}. Because of the relation between the carbon cycle on Earth and its tectonic activity, we note that a ``habsignature'' might also indicate the presence of plate tectonics on exoplanets.

In this context, we propose that observations of CO$_2$ absorption in exoplanetary spectra can be used as a signature for habitability (since $\rm CO_2$ can be removed from the atmosphere by dissolving into surface liquid water before being stored within the crust by tectonic activity) and also potentially as a biosignature (since biology traps carbon, for instance, as sediments and hydrocarbon deposits). 
The use of CO$_2$ depletion (``dCO$_2$'') as a habsignature provides a rapid empirical diagnostic about which terrestrial exoplanets are likely to be habitable (i.e. host surface liquid water). This approach also happens to identify planets that are most similar to the current Earth. The main concepts of this paper are represented schematically in Figure~\ref{fig:figure1}.

\section{Reframing the search for liquid water and signs of life}\label{signature}

Conventionally, the search for liquid water and for signs of biology rely on the detection of added observables to the planetary signal. Ocean glints \cite{Sagan1993} and hydrogen exospheres \cite{Kameda2017} are both added signals to the planet's. For life, observables are limited to atmospheric compounds viewed as ``life products'' that otherwise would not be thermodynamically stable within the atmosphere of a lifeless planet. Examples shown in Figure~\ref{fig:figure2} include the presence of a high concentration of $\rm O_2$, which itself can yield $\rm O_3$ \cite{Meadows2018,Catling2018} and ionospheric O$^+$ \cite{Mendillo2018}, the presence of $\rm CH_4$ within a $\rm CO_2$-dominated atmosphere \cite{Krissansen2018}, and the presence of other hydrocarbons (e.g., ethane \cite{Villanueva2011}, chlorofluorocarbons \cite{Pidhorodetska2020}, etc.) or other molecules (e.g., nitrogen compounds  \cite{HaqqMisra2022,Schwieterman202}). Here, we propose to reframe this search by detecting the depletion of an observable rather than the addition of one. By doing so, we provide a new framework to consider  the search for signs of liquid water and/or life which will result in new pathways for these quests, complementary to those presented previously in the literature, thereby improving the prospects of identifying habitable/inhabited worlds. We present one such new pathway below. 

One of the most striking differences between present-day Earth and its neighboring planets relates to its low atmospheric carbon mixing ratio. As shown in Figure~\ref{fig:figure2}, Earth's atmosphere has a remarkably low level of carbon dioxide among Solar System terrestrial planets. While the atmospheres of Venus and Mars are composed of $>$95\% CO$_2$, Earth's atmosphere only contains $\sim$0.04\% CO$_2$ \cite{vanThienen2007}. Most of the atmospheric CO$_2$ that was present on early Earth is now sequestered in its rocks \cite{Walker1985,Pierrehumbert2010,SHIBUYA2013}. Within an order of magnitude, there is a similar mass of carbon nowadays within Venus's atmosphere as there is locked as minerals within Earth's crust \cite{Walker1985,Lecuyer2000}. Earth's atmospheric carbon has been depleted  by its hydrosphere and its biosphere. This is why atmospheric carbon depletion in a temperate rocky planet is a tracer of water oceans (a habsignature), or biomass (a biosignature), or both at once, which we refer to as a ``habiosignature'' (Figure~\ref{fig:figure2}).
Transition periods such as the Archean, during which Earth was both habitable and inhabited, might be harder to diagnose with dCO$_2$. However, we note that Earth's atmosphere, whilst possessing a much larger carbon content than current day's, already had a depleted CO$_2$ level (ranging from $\sim 10\%$ at $4~\rm Gyr$ to $\sim2\%$ at $2.5~\rm Gyr$), with N$_2$ already being the dominant compound \cite{Catling2020}. 

\subsection{Atmospheric carbon depletion as a biosignature}

Biology --as we know it-- does not just produce chemicals, it also consumes them. Thus, biology cycles elements and removes some molecules from its environment, such as sulfur, nitrogen, and carbon-bearing molecules to make other products \cite{Sethi2019}. Here, we focus on carbon cycling and more specifically, on the removal of atmospheric CO$_2$, which is expected to be the primary carbon-bearing molecule in the atmosphere of terrestrial planets \cite{Gaillard2021}. The two main processes for biological carbon-dioxide depletion on Earth are (1) oxygenic photosynthesis, where carbon is sequestered in soils and hydrocarbon deposits, and (2) the creation of shells, made from calcite and aragonite ($\rm CaCO_3$), which are stored as sediments in rivers and oceans \cite{Archer1996,Bednarsek2012}. Accordingly, life on Earth plays a substantial role in the carbon cycle \cite{Archer2009} with about 20\% of modern global carbon sequestration being biologically-driven \cite{Plank2019}. In fact, biology's ability to extract carbon from the atmosphere may have initiated one of the {\it snowball Earth} events \cite{Kopp2005}. This being written, in most cases it is expected that biology plays a minor role in the sequestration of carbon, as its ability to fix carbon is ultimately capped by the water cycle within which it operates \cite{Plank2019}. Therefore, other signatures, such as O$_3$, need to be considered to fully ascertain the presence of biological activity when a substantial depletion of atmospheric carbon is detected. We address this point in Section \ref{observe}.

Similar arguments have been proposed to interpret a depletion of SO$_2$ \cite{Bains2021,Jordan2022} in the clouds of Venus, low concentrations of hydrogen compounds on Titan \cite{McKay2005,McKay2016}, and depleted CO within methane-rich atmospheres \cite{Kharecha2005,Schwieterman2019,Sauterey2020,Thompson2022}. Although controversial, a related example to a depletion as a signature in exoplanet literature is the concept of anti-biosignatures (e.g. large concentration of CO \cite{Gao2015,Wogan2020} or H$_2$ \cite{Catling2018,Hoehler2022}), where in that particular context the absence of a depletion is proposed as the signal.

\subsection{Atmospheric carbon depletion as a habsignature}

Liquid water readily dissolves atmospheric CO$_2$ \cite{DIAMOND2003,Mitchell2010} which can facilitate the formation of carbonates \cite{Zeebe2012}, creating a detectable habsignature \cite{Honing2021}. 
CO$_2$ is expected to dominate the speciation of outgassing products of {rocky planets with a bulk composition and redox state similar to Earth's, followed by CO, H$_2$O, and N$_2$ \cite{Gaillard2021}. When Earth cooled to temperatures which allowed liquid water, much of the atmospheric CO$_2$/CO dissolved into the new global ocean. Komatiite, a mafic to ultramafic igneous rock that paved the seafloor on early Earth, was exposed to this CO$_2$-enriched seawater and, as a result, underwent extensive mineral carbonation which caused an initial decrease in the CO$_2$-concentration of seawater and atmosphere \cite{SHIBUYA2013}. Evidence of mineral carbonation is not limited to Earth. Carbonate-alteration of the Martian meteorites Lafayette and ALH84001 indicates that mineral carbonation likely occurred during the Noachian and Amazonian periods when liquid water was available on Mars's surface \cite{Tomkinson2013,Steele2022}. There are also signs of mineral carbonation on Ceres \cite{Milliken2009}. Therefore, similar processes can be expected to operate on habitable rocky exoplanets \cite{Graham2020,Hakim2021}. 

The speciation of gases during magmatic degassing is strongly influenced by atmospheric pressure \cite{Gaillard2021}. For planets with similar bulk composition as Earth, while  water may dominate the mix for pressures under $\sim$3 bars, above that threshold, carbon dioxide is expected to dominate. Current models therefore suggest that temperate terrestrial planets are able to produce a large (up to $\sim$100 bars) CO$_2$ atmosphere via outgassing \cite{Wordsworth2013,Gaillard2021} and, if their surface temperature allows for water condensation, to sequester a large faction of it in their hydrosphere. If terrestrial exoplanets that underwent igneous differentiation expose Mg- and Fe-rich (mafic or ultramafic) lithologies to CO$_2$-rich fluids at temperatures lower than $300^\circ\rm C$ \cite{Klein2011}, carbonate minerals can form at the expense of primary igneous minerals, such as olivine, which is the most abundant mineral in our Solar System. This process, referred to as mineral carbonation, is rapid on geological timescales \cite{Kelemen2008}. 

It is possible that Venus was more similar to Earth during most of its history than it is presently. Current observations are compatible with Venus being both tectonically active and hosting habitable conditions on its surface as recently as $1$ to $0.7~\rm Gyr$ ago \cite{Way2016,Krissansen2021}. If this was the case, rising temperatures could have lead its surface liquid water bodies to evaporate and its hydrogen to escape to space, as evidenced by its D/H ratio \cite{Kasting1984}. Since outgassed CO$_2$ dominates Venus' atmosphere, there would not have been an efficient carbonation process in place since. Observations are also compatible with Venus being never able to condense its water vapor into oceans \cite{Hamano2013,Turbet2021b,Krissansen2021}.
In either scenario, relatively low carbon abundance in the atmosphere of a temperate rocky exoplanet, compared to other planets within the same system, appears as a reliable habsignature of recent and extensive amounts of liquid water at its surface.


\subsection{Atmospheric carbon depletion as an habiosignature}

Atmospheric carbon depletion is a habiosignature because it can indicate the presence of a hydrosphere and/or of a biosphere, both planet-shaping processes of interest to astronomers. 
A planet's concentration of CO$_2$ has long been known to be important to maintain its habitability. Research so far has focused on determining CO$_2$ concentrations which permit water to remain in a liquid phase at the surface, \cite{Bean2017,Turbet2019,Lehmer2020,Graham2020,Krissansen2022} in particular under the assumption of an Earth-like carbon cycle.  
For example, Refs. \cite{Bean2017,Lehmer2020} propose empirical tests to the concept of the habitable zone, by making the hypothesis of having a population of terrestrial planet large enough to be able to statistically study how the CO$_2$ mixing ratio evolves with planet’s insolations. Refs. \cite{Turbet2019,Graham2020} propose additional empirical tests by statistically studying how the CO$_2$ mixing ratio evolves between planets in and out of the Habitable Zone. Here, we take a complementary approach by considering that habitable planets will often have non-habitable neighbors.
We argue that a terrestrial planet with an atmospheric carbon abundance orders of magnitude lower than others within the same system implies the presence of a hydrosphere and/or of a biosphere. Moreover, the water and carbon cycles are intertwined in a biological and geological context. Therefore, the concept of ``habiosignature'' might go further than semantics since {\it Gaian cycles}, a complex interplay of geological and biological processes \cite{Lovelock1974,Lenton2016,Arthur2023}, are likely responsible for the relative long-term climate stability of our planet \cite{Sagan1972,Walker1981,Lenton2016,Arnscheidt2022}, of which atmospheric carbon depletion is a tracer.

Whilst we argue that a depletion in atmospheric carbon is a habiosignature, we now focus more specifically on CO$_2$ (and thus a depletion in CO$_2$, dCO$_2$) for most of the remainder of the paper, primarily because of the ease of measuring its spectral features in the mid-infrared, and its applicability to rocky planets with compositions similar to Earth's.

\subsection{Practical applications of detecting dCO$_2$.}
Systems with multiple terrestrial planets (with atmospheres) are needed before attempting to use dCO$_2$ as a hab- and habiosignature. As little as one additional terrestrial planet within the system might be sufficient. Here we stress that the more planets in a given system, and the closer they are to the planet being investigated, the better the calibration and thus the stronger the reliability of a depleted feature such as dCO$_2$. The physical properties of exoplanetary systems have low entropies, meaning that planetary systems are expected and found in ``peas-in-a-pod'' architectures \cite{Weiss2018,Sandford2021,Mishra2021,Millholland2021,Goyal2022}. Similarities between neighboring planets helps to calibrate depletions for a given planet. Proximity to the snow line could affect a planet's C/O ratio \cite{Oberg2011}; however, this transition is not expected to lead to orders-of-magnitude differences in the initial carbon inventory, and  C/O can be estimated from the atmospheric composition \cite{Madhusudhan2019,Ahrer2023}. 

Observations using direct imaging (e.g. using METIS with the ELT \cite{Snellen2015}) would be the most convenient. Compared to the transit method, it is more likely to identify all planets of a system, including those that might have a different redox state (having formed further from their star). Direct imaging is also a practical route to avoid a number of false positives and false negatives that we outline in Section~\ref{false}. Unfortunately, the study of temperate rocky worlds via directing imaging remains out of reach to this day, whereas such planets are now accessible via transmission spectroscopy.

\subsection{Detectability of the dCO$_2$ habiosignature with JWST.}

Currently, JWST can search for dCO$_2$ for terrestrial planets transiting late-M dwarfs, which are expected to be abundant \cite{Dressing2015,Gaidos2016}.
Figure~\ref{fig:figure3} shows that detecting the presence of an atmospheric carbon depletion in the atmosphere of temperate terrestrial planet with JWST is within reach. Specifically, it shows that for a planet like TRAPPIST-1~f the mixing ratio of CO$_2$ can be constrained to within $\sim$0.5~dex within 40 transits (Figure \ref{fig:figure3}, middle). This results holds for planets TRAPPIST-1~e, g, and h, as they are expected to have a similar SNR to planet f, while the three innermost planets would require less than half that number of transits \cite{Gillon2017b,Lustig2019}. 
While elemental composition varies between different stellar systems (and between rocky planets from different star systems \cite{Hinkel2014}), within the same system planetary elemental composition are expected to be roughly similar \cite{Oberg2011,Putirka2019}. This results in a capability to distinguish between high- and medium-depletion levels (respectively X$_{\rm CO_2} = 400~\rm p.p.m.$ and X$_{\rm CO_2} = 30\%$) at the $\sim$10$\sigma$ level. It also allows to distinguish between medium- and negligible-depletion levels (respectively X$_{\rm CO_2} = 10\%$ and X$_{\rm CO_2} = 95\%$) at the $\sim5\sigma$ level.

\section{Three birds, one stone: observations of the $4.3~\rm \mu m$ CO$_2$ band}\label{observe}

In addition to possibly yielding the detection of an habiosignature, the spectral range surrounding CO$_2$'s strong $4.3~\rm \mu m$ absorption band \cite{Gordon2022} in the NIR offers observational benefits that support two other characterization steps: (1) the detection of atmospheres for terrestrial exoplanets, and (2) the distinction between a biological and geological atmospheric carbon depletion. 

\subsection{Detecting the atmosphere of a temperate terrestrial exoplanet}

To date, it has not yet been possible to reveal the presence of an atmosphere around a temperate terrestrial exoplanet, which is the very first step of atmospheric characterization. Up until now, it was only possible to search for hydrogen-dominated atmospheres using the {\it Hubble} Space Telescope \cite{dewit2016,dewit2018, Wakeford2019, Garcia2022,Gressier2022}. Precise measurements of planetary physical parameters, notably transiting timing variations, can also constrain the presence/absence of a large hydrogen-dominated atmosphere \cite{Turbet2020,Agol2021}. JWST and the upcoming ELTs now offer the possibility to search for the presence of atmospheres for favorable terrestrial targets, such as the TRAPPIST-1 planets, in as little as 5 to 10 transits \cite{Lustig2019,Fauchez2019}. More transits, however, may be needed due to (1) stellar contamination of the planets' transmission spectra \cite{rackham2017, rackham2018} and/or (2) the presence of clouds/hazes \cite{Fauchez2019,Turbet2020,Komacek2020,May2021}.

We show in Figure~\ref{fig:figure4} that there exists a sweet spot where the combination of both effects and the photon noise is optimal. The effect of stellar contamination decreases while photon noise increases beyond $\sim1~\rm \mu m$ for stars like TRAPPIST-1.  Therefore, a local minimum exists, dependent on the number of transits needed (photon noise). Figure~\ref{fig:figure4} presents the uncertainty associated with the transmission spectrum of a TRAPPIST-1 planet gathered over 10 transits --the nominal amount for the reconnaissance of atmospheres around temperate terrestrial planets. To allow for an easy comparison we report the uncertainty at a uniform spectral resolving power of $R\sim30$. The local minimum for such an exploration program is found between $4$ and $10.5 ~\rm \mu m$. This wavelength range overlaps with the wavelength range over which hazes have a marginal impact on a planet's transmission spectrum, namely wavelengths above $\sim 3.3~\rm \mu m$  \cite{Fauchez2019} although the exact threshold depends on the aerosol size distribution.

The precision sweet spot for the reconnaissance of terrestrial planets with JWST via transmission spectroscopy thus lies between $4$ and $10.5~\rm \mu m$. In this spectral window, one absorption feature dominates across all models of secondary atmospheres contemplated in the literature: the CO$_2$ $4.3~\rm \mu m$ band \cite{Fauchez2019,Turbet2020}. In fact, as shown in Figure~\ref{fig:figure1}, this band's absorption typically dominates absorption by molecules at shorter wavelengths as well. 

CO$_2$ absorption bands are of particular interest for atmospheric reconnaissance (i.e., a low SNR spectrum) as they are both strong and sharp. The bands are strong because CO$_2$ is an excellent absorber in the infrared \cite{Gordon2022}, even at a low atmospheric concentration (like in Earth's atmosphere). In fact, this molecular feature is the first reported by JWST in the atmosphere of an exoplanet \cite{JWSTTECERS2023, Alderson2023, Rustamkulov2023}. CO$_2$'s bands are sharp and dense in opposition to, e.g., H$_2$O's and CH$_4$'s broad and ``sparse'' absorption bands \cite{Gordon2022}. These two qualities of the CO$_2$ bands allow them to be more prominent in low-resolution spectra and thus not drowned out by correlated noise. As a result, the $4.3~\rm \mu m$ CO$_2$ band appears to be an optimal signature to detect the presence of secondary atmospheres around terrestrial planets, whilst being accessible to JWST and has been proposed for JWST observations already as a diagnostic test of the presence of a terrestrial planet atmospheres \cite{Stevenson2021}.

\subsection{Habitable or inhabited?}

Once an atmosphere has been detected around a temperate terrestrial exoplanet, we can reasonably expect that an extensive observing campaign will be dedicated to studying its atmospheric properties more in detail. For context, the atmospheric reconnaissance of the TRAPPIST-1 system initiated during JWST's Cycle 1 has already been supported by eight different observing programs, totaling over 200 hours. The scale of such a large characterization program will be capped to $\sim$100 planetary transits, limited by the availability of windows of opportunity and by the telescope's lifetime. We use the {\sc Tierra} code \cite{Niraula2022} to model planetary spectra and show that $\sim40$ transits would be sufficient to yield constraints on the abundances of the dominant atmospheric absorbers.  Figure~\ref{fig:figure3} shows that the abundance of strong absorbers such as CO$_2$ can be constrained to within $0.5~\rm dex$, down to the $10~\rm p.p.m.$ level (Earth's CO$_2$ concentration, for comparison, is $400~\rm p.p.m.$), thereby allowing to assess the presence of a habiosignature around planets like TRAPPIST-1's. 

Assuming a habiosignature is revealed in the atmosphere of a temperate terrestrial planet, the next natural step will be to disentangle between a geological and biological origin, i.e., a habitable and inhabited environment. Is the observed dCO$_2$ driven mostly by a strong water cycle or also by biology? To answer this question, we evaluate the viability of two avenues that leverage the wavelength range surrounding the $4.3~\rm \mu m$ CO$_2$ band: (1) constraints on the isotopic fractionation of CO$_2$ and (2) constraints on the abundance of O$_3$. 

\subsection{Insights from isotopic fractionation}

Isotopic fractionation can be remotely detected \cite{Molliere2019,Line2021}, owing to the fact that a change in a molecule’s mass distribution results in shifts in its absorption lines. For CO$_2$, the main difference in the $4.3~\rm \mu m$ band between the dominant isotopes ($^{12}$C$^{16}$O$_2$ and $^{13}$C$^{16}$O$_2$) is a substantial shift of the bandhead by $\sim~0.2~\rm \mu m$ (Figure~\ref{fig:figure5}, left). 

Isotopic fractionation is notably driven by atmospheric escape, biological activity, and water-rock interactions. In general, biological activity preferentially incorporates the light isotopes (e.g., $^{12}$C and $^1$H) into organic compounds; the removal of those organic compounds then drives the residual carbon toward heavier values. Plants and algae have approximate $\delta^{13}$C values between $-10$ and $-40 \times 10^{-3}$ \cite{Farquhar1989} and a $\delta^{2}$H between $-90$ and $-180 \times 10^{-3}$ \cite{Estep1980}. 

Despite the presence of a global biology on Earth that contributes to about 20\% of global carbon sequestration \cite{Plank2019}, Earth’s atmospheric carbon isotopic fractionation is consistent with other objects in the Solar System \cite{Woods2009}. Similarly, Earth’s rock record shows no substantial changes in $^{13}$C/$^{12}$C over the last 3.5 billion years \cite{Schidlowski2001,Krissansen2015}, challenging our current understanding of isotope fractionation in extant metabolic pathways \cite{Garcia2021}.

The picture further complexifies when one accounts for other sources of isotope fractionation that also favor the removal of light isotopes, such as atmospheric escape \cite{Jakosky2017,Jakosky2018}. In addition to interpretation challenges, the detection of carbon fractionation in the atmosphere of a terrestrial planet will likely be out of reach with upcoming observatories. Figure~\ref{fig:figure5} shows the expected contribution of $^{13}$CO$_2$ to be around the $5~\rm  p.p.m.$ level for an isotopic fraction similar to Earth's atmosphere, and of $\sim$$15~\rm  p.p.m.$ level for $100 \times$ Earth’s. These values are smaller than the $\sim20~\rm p.p.m.$ precision expected on such a spectral bin with $\sim$100 transit observations (Figure~\ref{fig:figure5}, right panel). Therefore, carbon isotopologues are unlikely to be informative biosignatures for the JWST era \cite{Glidden2022}.

\subsection{Insights from oxygen abundance}

While the sequestration of atmospheric CO$_2$ by a planet's hydrosphere allows for the removal of both carbon and oxygen \cite{DIAMOND2003,Mitchell2010,Zeebe2012}, the sequestration of atmospheric CO$_2$ by a planet's biosphere leads primarily to the removal of just the carbon, leaving oxygen as a by-product \cite{Sagan1993,Meadows2018}. For redox states similar to Earth's, CO$_2$ is expected to be the dominant molecule in the secondary atmospheres of terrestrial planets by orders of magnitude \cite{Gaillard2021}. If CO$_2$ is depleted in a substantial manner by a photosynthetic biosphere, a large fraction of O$_2$ will consequently be returned to the atmosphere. 

While O$_2$ is challenging to detect directly for known transiting exoplanets with upcoming observatories \cite{Fauchez2020,Rodler2014}, its presence can be inferred by detecting its photochemical byproduct O$_3$ \cite{Ratner1972,Kasting1980,Leger1993,Meadows2018}. Indeed, O$_3$ offers a series of strong absorption bands, similar to CO$_2$, including one at $4.8~\rm \mu m$ \cite{Gordon2022} in the vicinity of CO$_2$'s $4.3~\rm \mu m$ band. An Earth-equivalent level of ozone can readily be detected at the $\sim$5-$\sigma$ level with $\sim$100 transits of TRAPPIST-1~f observed with JWST/NIRSpec (Fig. 5, right panel). This means that ozone concentrations down to the $10~\rm p.p.m.$ level will be detectable for temperate terrestrial exoplanets with JWST \cite{Barstow2016}. We note that Earth's ozone level ($10$--$15~\rm p.p.m.$) may serve as a lower limit considering the higher level of stellar UV irradiation that temperate terrestrial exoplanets orbiting late M dwarfs will face, as they are the only terrestrial exoplanets amenable for atmospheric characterization with JWST \cite{Triaud2021}.

O$_3$ could also be produced via the photodissociation of CO$_2$ into CO and O$_2$. However, only a small fraction of CO$_2$ is expected to be dissociated in such a way as seen for Venus and Mars (see next section) leaving an even smaller fraction of the resulting O$_2$ turned into O$_3$. Thus, while the atmospheric depletion of CO$_2$ is an habiosignature preventing from distinguishing between the presence of a hydrosphere and a biosphere and that O$_3$ is an ambiguous biosignature just like O$_2$, together, they form a reliable and readily accessible biomaker for the JWST era.

\section{False positives \& false negatives}\label{false}

Any signature has false positives and false negatives that can seriously impact the interpretation of a detected feature \cite{Meadows2018,Catling2018}, or in the present case, the identification of a depleted feature. However, we find that a depleted CO$_2$  (dCO$_2$) as a habiosignature has few such issues. 

\subsection{False positive scenarios} A false positive to dCO$_2$ as a habiosignature would imply the detection of an atmosphere depleted in CO$_2$ without the contribution of a hydro- or a biosphere. We consider five such possible scenarios in the following.

\begin{itemize}
\item \textbf{Dry Sequestration.} Some rocks have a demonstrated ability to react with atmospheric CO$_2$ and sequester it such as brucite, peridotite or serpentinite \cite{Kelemen2008}, which are considered for engineered carbon capture and sequestration to combat climate change. However, mineral carbonation of rocks is relatively inefficient when CO$_2$ is dry \cite{WANG2019}, or if temperatures are low ($\ll100^\circ~\rm C$) \cite{Kelemen2008}. Reactions leading to carbon sequestration usually need access to water and the kinetics of all known carbon mineralization reactions currently occurring on Earth are greatly enhanced by water \cite{KELEMEN2020}. For instance, brucite is one of most reactive minerals in the presence of CO$_2$ as long as water is available; brucite remains unreactive under a dry CO$_2$ atmosphere \cite{Loring2012}.  The deep carbon cycle between the exosphere is critical in controlling atmospheric levels of CO$_2$ and the solid Earth operates on timescales $>10^5~\rm years$ \cite{Zeebe2012}. However, the removal of CO$_2$ from the atmosphere is closely linked to the presence of water, either directly through the uptake of CO$_2$ in the surface ocean or indirectly via the dissolution of silicate minerals and precipitation of carbonate minerals. Using the Solar system as an example, even at high CO$_2$ fugacities, like on Venus, and high surface temperatures, CO$_2$ is stable in the atmosphere and demonstrates that dry carbonation of rocks has a likely minimal impact on carbon removal and is an unlikely false positive under dry conditions (Venus has lost most of its water \cite{Kasting1993}).

\item \textbf{Nightside Cold Trap.}  In a scenario of atmospheric collapse, most of the atmospheric CO$_2$ freezes on the night-side of a planet (tidally-locked or not) \cite{Heng2012}, leaving low atmospheric CO$_2$ partial pressures in some cases \cite{Turbet2018}. Atmospheric collapse gets increasingly likely for planets further from their star. A collapse can be identified in some cases by measuring an exoplanet's atmospheric temperature using transmission spectra. In addition, planets with measured atmospheric compositions compatible with this situation will need detailed and dedicated 3D numerical climate simulations to contextualize their CO$_2$ concentrations before any statement is made regarding their habitability. All the TRAPPIST-1 planets are expected to be tidally-locked \cite{Gillon2017b}. For planets like TRAPPIST-1f, that we used in our simulations, atmospheric collapse is not expected for pressures typically $> 0.7~\rm bar$, \cite{Turbet2018}.

\item \textbf{Photodissociation of CO$_2$.} In this scenario, CO$_2$ is photo-dissociated or chemically transformed into other compounds. Our premise, however, is about atmospheric carbon abundance and not about just CO$_2$ abundance. The most likely carbon-bearing product from an atmospheric reaction involving CO$_2$ is CO \cite{selsis2002}, which is also observable in the same $4-5~\rm \mu m$ band (Figure 5) on which we encourage efforts to focus \cite{Rustamkulov2023}. As such, it is possible to measure the overall atmospheric carbon abundance with the same data. In any case, it is expected that at most $10\%$ of CO$_2$ would transform into CO and O$_2$ \cite{selsis2002} for an atmosphere initially containing $1~\rm bar$ of CO$_2$, which would be detectable with JWST (Figure 5). Atmospheres with higher partial pressures in initial CO$_2$ produce fractionally less CO \cite{James2018} as seen for instance on Venus. Furthermore, we advocate measuring and comparing several planets of the same system together since photo-dissociative effects are expected to decrease with orbital separation. 
Photo-dissociation is expected to decrease quadratically with distance from the star, in a monotonic function. As such we do not expect order-of-magnitude differences between a series of adjacent planets, allowing a means to diagnose this process from others. The cases of Venus and Mars highlight that this is not a scenario of particular concern \cite{selsis2002}.

\item \textbf{Photodissociation of other species} Another aspect to consider relates to the production of ozone as a product of the photodissociation of stratospheric water. Refs. \cite{Kopparapu2017,Badhan2019} have shown that slow-rotating terrestrial planets orbiting M dwarfs can have moist greenhouse conditions with significant levels of water vapor in their stratospheres. Late-M dwarfs such as TRAPPIST-1 are active longer into their life cycles and have stronger far-UV and weaker near-UV emissions, leading to exotic photochemistry \cite{Segura2005,Gao2015,Rugheimer2015,Badhan2019,Wunderlich2019}. This suggests that stratospheric water may produce a false positive detection of the biosignature dCO$_2+\rm O_3$.


\item \textbf{Reduced Interior.} A planet's atmospheric composition depends on whether its interior is oxidized or reduced \cite{Liggins2022,Lichtenberg2022}. In the latter case, instead of CO$_2$, CO and CH$_4$ are expected to be the main carbon carriers. Neither CO nor CH$_4$ dissolves well in liquid water, preventing carbon depletion via the hydrosphere. 
Planetary bulk compositions are expected to be increasingly reduced with orbital distance from their host star \cite{Wordsworth2018}, something that is also seen in the Solar system. Planets with different bulk compositions are also expected to outgas a different atmospheric chemistry \cite{Gaillard2021}. The detection of CO$_2$ in large concentrations for an outer planet indicates an oxidised world and would calibrate the inner system, and allow the detections of depleted feature on those planets.
Disc-driven migration is known to move planets from their birth place, and planetary objects similar to Titan, Triton and Pluto, assembled in Hydrogen-rich parts of the disc might be brought to the habitable zone \cite{Oberg2011, Ormel2017, Bitsch2019,Venturini2020}. These objects have different bulk densities, and will be identified from their position in a mass/radius diagram \cite{Ortenzi2020}. In addition, should a temperate rocky planet be in a reduced state, this can be easily diagnosed from their atmospheric composition \cite{Ortenzi2020,Bower2022}. CO would show a prominent feature in the same $4-5~\rm \mu m$ range with CO$_2$ absent (Figure~\ref{fig:figure5}, right panel). CH$_4$ has an absorption band at $3.3~\rm \mu m$, likely to be observed with the same instrumental setup. Both the CO and the CH$_4$ features present higher level of significance than CO$_2$'s. For reduced sets of planets, a different set of bio and habsignature would need to be found. 

\item \textbf{Other Surface Solvent.} In this scenario, an abundant liquid solvent other than liquid water is able to dissolve and sequester carbon efficiently at room temperature. CO$_2$ is soluble in other liquids such as ethanol (alcohols) or polyethylene glycol \cite{Aschenbrenner2010,DALMOLIN2006} but, to our knowledge, such compounds do not form abiotically in significant amounts on a planetary scale. Liquid methane is a fairly efficient solvent for CO$_2$ \cite{FOGG1992} that can form in large amounts on a planetary scale, for instance, Titan's lakes. However, liquid CH$_4$ is only stable at extremely low temperatures ($\sim 120~\rm K$) that are not habitable for life as we know it.  Water remains one of the most abundant molecules in the Universe, and the most abundant solvent expected to be outgassed and condensed for a temperate terrestrial planet \cite{Gaillard2021}. 
\end{itemize}

From their very first detections, exoplanets' orbital and physical properties have defied expectations \cite{Wolszczan1992,Mayor1995,Lovis2006,Doyle2011}. We have tried to explore many false positive signatures, but unknown and efficient CO$_2$ removal processes could well operate on an exoplanet. In this event, the detection of an atmospheric carbon depletion would not be attributed to the presence of a hydrosphere or a biosphere, but it would reveal an unexpected sequestration mechanism efficient on a planetary scale. This depleted feature would thus still be of very high scientific interest and could possibly spark novel industrial exploration of processes capable of balancing anthropogenic emissions of CO$_2$, and address climate change on Earth. More research is needed to understand how atmospheric carbon reacts with different solvents, at different temperatures and pressures, how photochemistry in terrestrial planet atmospheres around M dwarfs work while considering different bulk compositions, as well as to study the possible nature and availability of such solvent alternatives.

\subsection{False negative scenarios} A false negative is a planet that possesses large amounts of liquid water and/or a biosphere, but its atmosphere is not depleted in CO$_2$. We consider three such possible scenarios in the following.

\begin{itemize}
\item \textbf{Subsurface Biosphere.} An atmosphereless planet can be habitable in its interior if habitability is defined to also include sub-surface liquid water. Such a situation is similar to Europa and Enceladus \cite{Reynolds1983,Vance2018}. In this particular case, this is an issue of definition. An atmospheric signature is only valid where there is a detectable atmosphere. Other planets might present low atmospheric pressures that are hard to detect, similar to the case of Mars. Those cases will be noticed easily. From their already known radius and/or the mass it is possible to deduce a past atmospheric escape.

\item \textbf{Saturated Oceans.}  Depleted CO$_2$ as a habsignature only functions if carbon continuously dissolves into the ocean. Plate tectonics moves the crust about and, in doing so, moves carbonates outside of oceans onto land, which we can see nowadays as limestone deposits. Tectonics also mostly buries carbonates out of reach of the atmosphere into the crust and back into the mantle \cite{SHIBUYA2013,Southam2015,Plank2019}. 
A false negative could emerge for rocky exoplanets without plate tectonics, and/or where the oceans become saturated and can no longer absorb any more CO$_2$ \cite{Kitzmann2015,Honing2021,Graham2022}. Without other known efficient sinks, CO$_2$ would therefore remain in the atmosphere. Should this scenario happen, other widely used biosignatures can still be employed, and used in exactly the same way as they are being considered nowadays. However, whilst the onset of plate tectonics is still a highly debated topic \cite{Bercovici2003,Korenaga2013}, it is generally accepted that liquid water is required to weaken the protoplates' yield stress \cite{Mei2000a,Mei2000b}, and that liquid water and carbonates lubricate the plates' motion. This means that tectonics might be an inevitability of a habitable/inhabited world \cite{Lovelock1974,Kasting1993,Lenton2016}. For packed, close-in systems of exoplanets, such as the TRAPPIST-1 system \cite{Gillon2017b}, plate tectonics might also be initiated tidally \cite{Zanazzi2019}, or via impacts \cite{Kral2018,Borgeat2022}. 

\item \textbf{Transient CO$_2$ concentration.} In this scenario, an extensive liquid water ocean is present but has not had time yet to significantly remove enough atmospheric carbon for a detection. The Archean Earth represents such a transition period. Yet, as discussed in Section~\ref{signature}, dCO$_2$ was detectable as a habsignature during most of the eon (Fig.~\ref{fig:figure3}). However as biomarker, it would likely be inconclusive since Earth was dominated at the time by methanogens \cite{Krissansen2018, Catling2020}, as we consider oxic biospheres (like the modern Earth) in this study. 
Earth's oceans contain about $2~\rm bars$ of CO$_2$ currently \cite{Lecuyer2000,Pierrehumbert2010}. If released in the atmosphere, the oceans would absorb those $2~\rm bars$ of CO$_2$ in $\sim 0.2~\rm Gyr$, at current absorption rates \cite{Catling2020,Plank2019,Zeebe2012}. Indeed, it is thought that most of Earth's primordial atmospheric CO$_2$ was dissolved by the time biological processes emerged on Earth circa $0.5~\rm Gyr$ after formation \cite{SHIBUYA2013,Catling2020}. This makes this false negative scenario likely to be observed in less than 1 in 50  habitable rocky planets. We note here that this scenario might be a faux- false negative, as a sudden release of bars of CO$_2$ (that could equally be produced by other events such as asteroid/comet impacts, or extensive volcanic episodes) would trigger a substantial temperature increase due to greenhouse effects \cite{Bean2017,Turbet2019}, thus reducing the habitability and inhabited nature of the planet temporarily.

\item \textbf{Habitability near the outer edge.} A planet near the outer edge of the habitable zone requires an important CO$_2$ partial pressure to maintain habitable conditions at the surface \cite{Kasting1993,Bean2017,Turbet2019,Lehmer2020,Krissansen2022}. For instance, moving Earth to the orbit of Mars requires about 2 bars of CO$_2$ to keep water in a liquid state. Our simulations (Fig.~3) show such a scenario is difficult to distinguish. After collecting 40 transits with JWST, any concentrations $>50\%$ CO$_2$ do not produce spectra significantly different ($<3\sigma$) than those from atmospheres with $100\%$ CO$_2$. Other habsignatures will need to be deployed for those cases.
\end{itemize}


\newpage

\section{Summary and future prospects}


Life on Earth is obvious. It shapes most aspects of our environment, including the composition of the atmosphere above us, and of the ocean and rocks below us \cite{Sagan1972,Lovelock1974}. Its continuity across several billion years is likely regulated by Gaian cycles, a complex system of geological, atmospheric and biological balances understood and nowadays investigated as ``Earth's systems'' \cite{Lovelock1974,Lenton2016,Arthur2023}. Life on Earth has a truly global effect. In other words, Life on Earth is planet-shaping. Planet-shaping Life is really what astronomers are after \cite{Arthur2023}. Should Life have arisen on Venus \cite{Greaves2021}, on Mars \cite{Krasnopolsky2004}, Europa \cite{Kargel2000} or Enceladus \cite{McKay2008}, whilst extremely fascinating, it is clearly not as planet-shaping, simply because it is not nearly as obvious in affecting what can be observed and measured about these celestial objects. Such Life would not be detectable on exoplanets, whereas planet-shaping Life is.

In this perspective, we outline how atmospheric carbon depletion, particular a carbon dioxide depletion, is a logical tracer of planet-shaping Life, and is able to reveal its global impact using remote-sensing methods of observation. An atmosphere such as Earth's becomes depleted in carbon thanks to the action of extensive amounts of surface liquid water, and/or by intense biological processes. Plate tectonics buries carbon away from the atmosphere, causing atmospheric depletion of CO$_2$ over geological timescales \cite{Southam2015}. 

In much of the scientific literature, observables are appreciated as a signal, an addition to something otherwise presumed pristine. For exoplanets, tracers of habitability --habsignatures-- such as the glint of a surface ocean \cite{Sagan1993}, and tracers of biology --biosignatures-- such as O$_2$ in quantities beyond those expected by photodissociation and chemical equilibrium \cite{Sagan1972,Meadows2018,Catling2018}, are all added signals. However, planet-shaping Life does not just produce but also consumes \cite{Sethi2019}, and the Earth's systems that sustain it (e.g. liquid water) also actively remove chemical species from the atmosphere \cite{DIAMOND2003, Mitchell2010,Zeebe2012}. We believe it is just as important to consider depleted signals as habsignatures and biosignatures. In the context of atmospheric carbon depletion, a depleted CO$_2$ feature compared to other planets within the same system is a good habiosignature, as it traces both habitable and inhabited environments. We explored the likelihood of false positive found all those we could imagine were unlikely to exist. We also investigated false negative signals, and propose means of addressing them. This is why we assess that CO$_2$ depletion is a robust habiosignature applicable to most temperate rocky exoplanets, particularly those most similar to Earth.

Measuring the CO$_2$ absorption feature comes with multiple benefits. Even at low concentrations, CO$_2$ absorption is the most detectable of all atmospheric features for a transiting temperate terrestrial exoplanet and can reveal whether the planet has an atmosphere, the first step in the characterization of any exohabitats. Furthermore, CO$_2$ is located at wavelengths where contamination from clouds and hazes is minimal, and where stellar contamination decreases below the typical photon noise, even for highly active late- M dwarfs. At these wavelengths, other molecules of interest have features too: CH$_4$, CO, and O$_3$, all of which are relevant in assessing the carbon abundance of the atmosphere, and in the case of O$_3$, in distinguishing whether carbon depletion is produced only by an ocean, or also by biology. 

These facts lead us to propose an observational strategy: to target the CO$_2$ feature, first with a few planetary transits, in order to assess the presence of an atmosphere and obtain an initial estimate of CO$_2$ concentration on several planets of the same system. Should these reconnaissance observations reveal depleted CO$_2$ features, these observations need to be augmented with a more extensive observing campaign, that will accurately measure their CO$_2$ abundances, but also the presence and concentrations of CH$_4$, CO, and O$_3$.

We sincerely hope this strategy is implemented by the multiple teams around the world that are attempting to explore the atmospheres of transiting temperate rocky exoplanets using space-based observatories like  JWST, but also planning to use  ground-based facilities such as the upcoming Extremely Large Telescopes. We also encourage all teams to investigate other depleted molecular features and bring about new tracers of habitability and biological activity. The more we have, the more reliable our interpretations of temperate rocky exoplanet spectra will be.

\vspace{0.5cm}

\newpage
\pagebreak
\clearpage

\section{Captions of Figures}

\begin{figure*}[h!]
\centering
\includegraphics[trim={0cm 0cm 0cm 0cm},clip, angle=0, width=0.95\textwidth ]{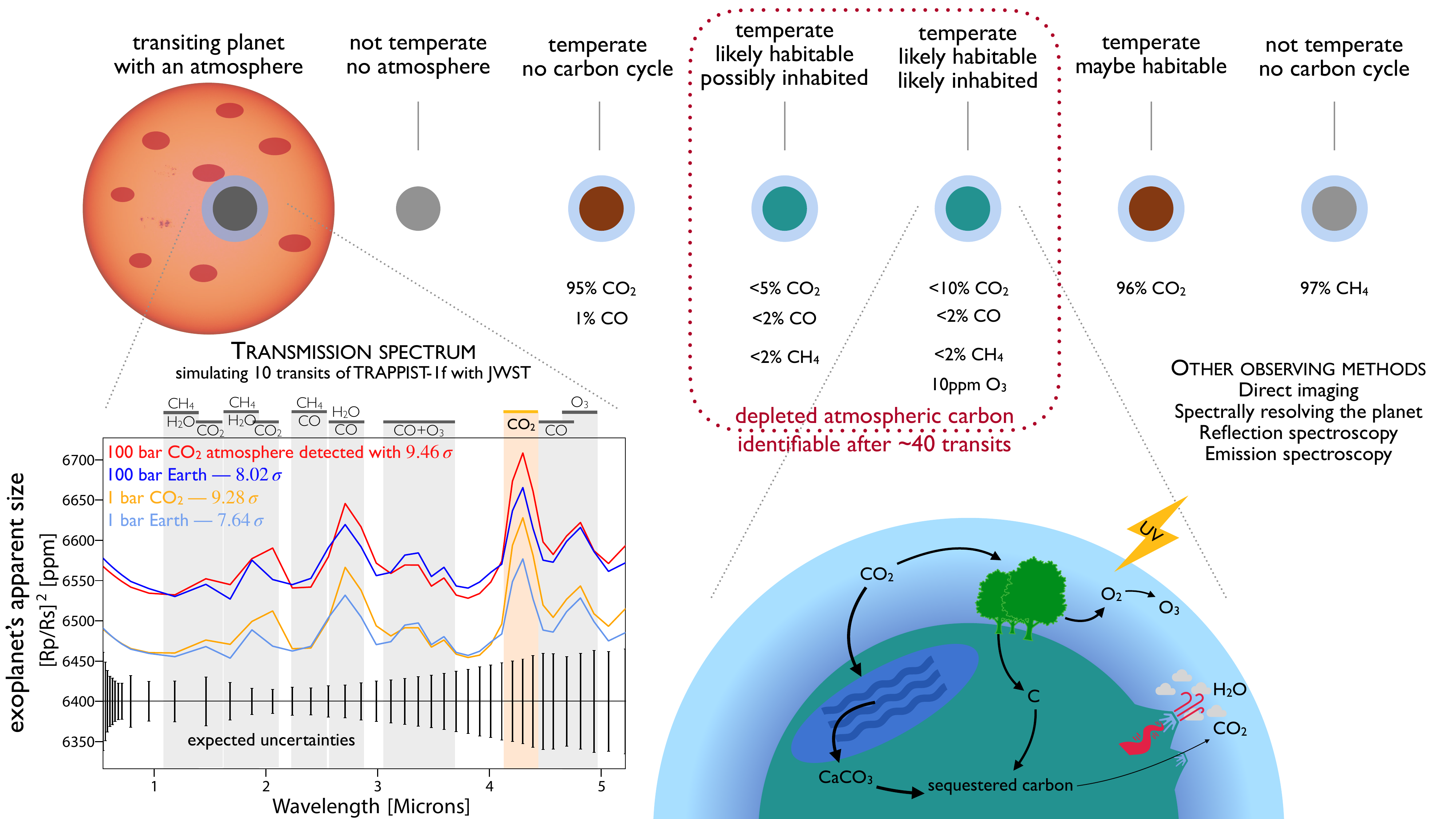}
\caption{Illustration of our strategy to detect habitable exoplanetary environment via CO$_2$ depletion. Each of the planets to the right hand side of the star describes a different scenario  discussed in the text (along with illustrative atmospheric concentrations). The panel on the left hand side depicts a simulation of the transmission spectrum of the temperate terrestrial planet TRAPPIST-1\,f. We explore the detectability of an atmosphere with $\sim$ 10 JWST/NIRSpec Prism transit observations, the minimum needed to produce a reliable diagnostic. Note that the deviation from a flat signal (no atmosphere) is primarily supported by the strong absorption features of CO$_2$, notably at $4.3~\rm \mu m$ (highlighted).  Exo-atmospheric models from \cite{Niraula2022} were used to make that panel. On the lower right hand side, we illustrate a simplified view of a carbon cycle involving surface liquid water and biology sequestering cycle and producing a depletion in atmospheric carbon. Atmospheric concentrations given near the planets are illustrative. 
\label{fig:figure1}}
\end{figure*}

\begin{figure*}[h!]
\centering
\hspace{-0cm}{\includegraphics[trim={0cm 0cm 0cm 0cm},clip, angle=0, width=\textwidth ]{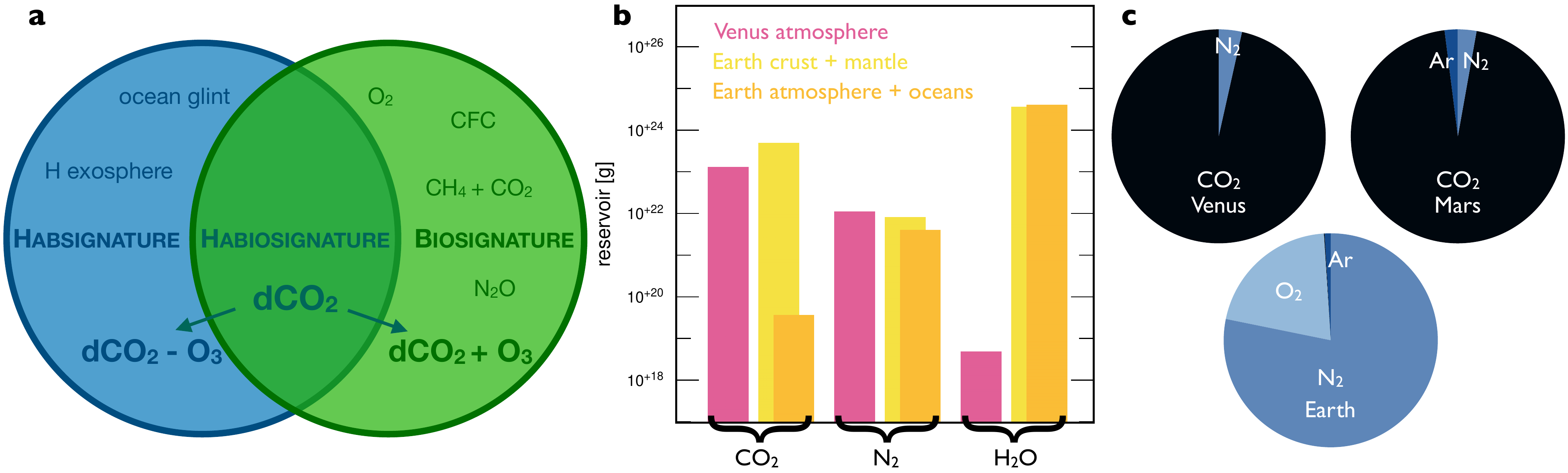}}
\caption{CO$_2$ depletion (``dCO$_2$'') as a signature of liquid water and/or life. Left: Schematic representation of observables associated with the presence of surface liquid water (``habsignature''), of a biomass (``biosignature''), or both (``habiosignature''). Centre: Comparison between present day Venus's, and Earth's interior and atmospheric volatile inventories (Data from Ref.\cite{vanThienen2007}). Right: relative atmospheric abundances for Venus, Earth and Mars.
\label{fig:figure2}}
\end{figure*}


\begin{figure*}[h!]
\hspace{-4cm}{\includegraphics[trim={0cm 0cm 0cm 0cm},clip, angle=0, width=1.5\textwidth ]{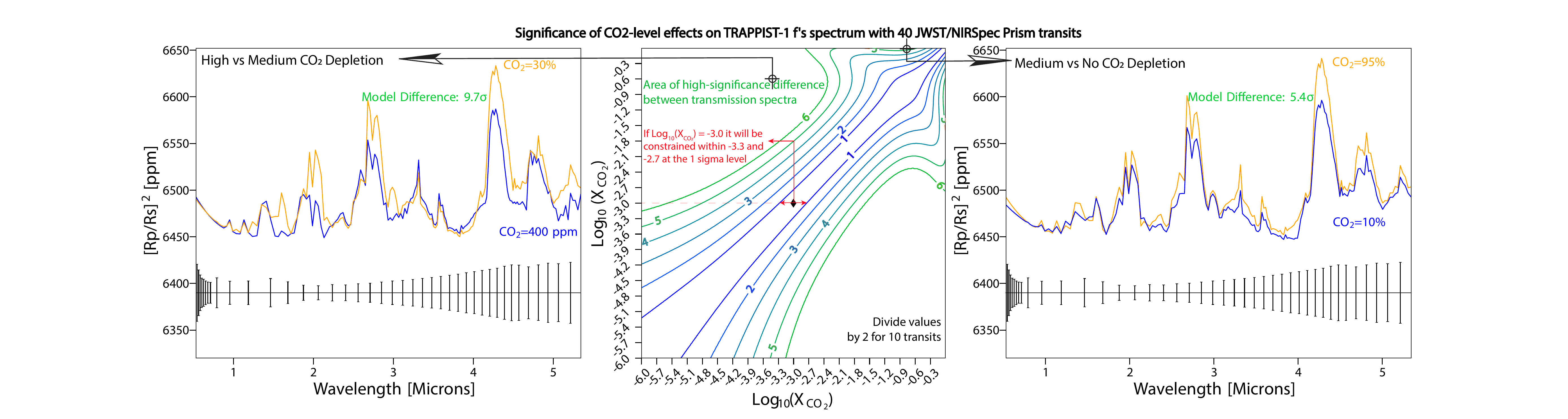}}
\caption{Detectability of CO$_2$ depletion (``dCO$_2$'') as an habiosignature with JWST. Middle: Significance map of CO$_2$-level effects on TRAPPIST-1~f's spectrum with 40 JWST/NIRSpec Prism transits. Values reported represented the deviation between two transmission spectra with different concentration of CO$_2$ (X$_{\rm CO_2}$). The region around the 1:1 line surrounded by the 1-$\sigma$ contour is at low-significance. The width of that region represents the confidence interval with which a given X$_{CO_2}$ can be retrieved (typically $\sim 0.5~\rm dex$, example provided in orange for Log$_{10}$ (X$_{\rm CO_2} = -3.0$)). Left: Comparison between the spectra of TRAPPIST-1~f with a high CO$_2$ depletion (X$_{\rm CO_2} = 400~\rm p.p.m.$) and medium one (X$_{\rm CO_2} = 30\%$). The spectra and their associated 1-$\sigma$ errorbar (black) are shown for a resolving power of 
$R\sim30$. Right: Comparison between the spectra of TRAPPIST-1~f with a medium CO$_2$ depletion (X$_{\rm CO_2} = 10\%$) and without any (X$_{\rm CO_2} = 95\%$). 
\label{fig:figure3}}
\end{figure*}


\begin{figure*}[h!]
\centering
\includegraphics[width=\textwidth, height=\textheight, keepaspectratio]{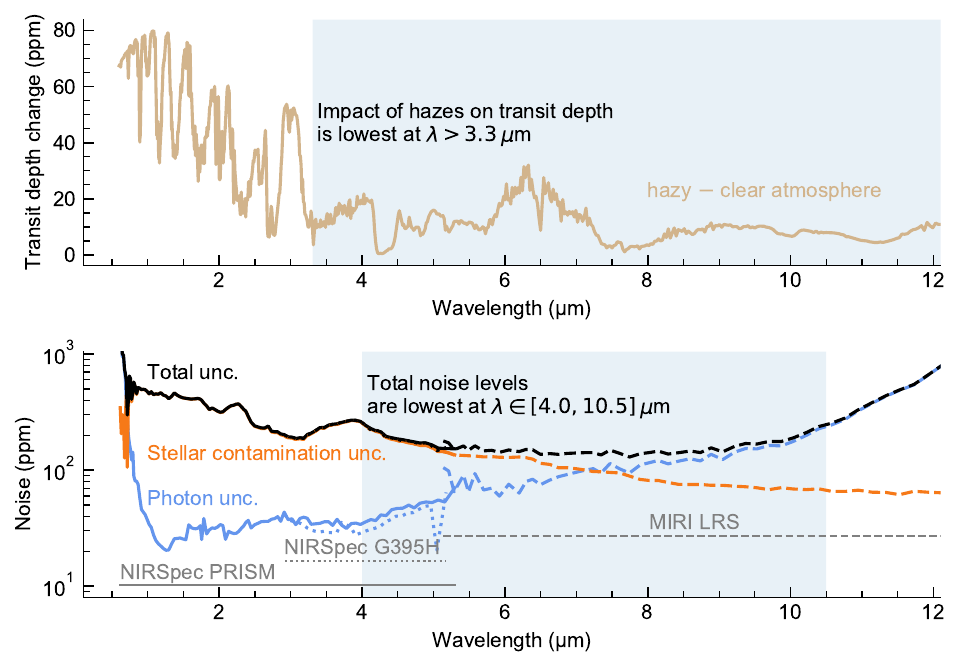}
\caption{
    Top: The difference between ``hazy'' and ``clear atmosphere'' transmission spectra for TRAPPIST-1e \cite{Fauchez2019}. The impact of haze on the transit depth is smallest at wavelengths greater than 3.3 $\mu$m.
    Bottom: Photon-limited uncertainties increase at the shortest and longest wavelengths, where the photon flux from the star is limited.
    Systematic uncertainty due to stellar contamination increases toward shorter wavelengths, where contrasts between spots and faculae with respect to the photosphere are greatest.
    The combined uncertainty from these noise sources reaches a minimum between 4--10.5\,$\mu$m.
    See details in Code Availability section.
    \label{fig:figure4}
}
\end{figure*}


\begin{figure*}[h!]
\hspace{-1cm}\includegraphics[trim={0cm 0cm 0cm 0cm},clip, angle=0, width=1.1\textwidth ]{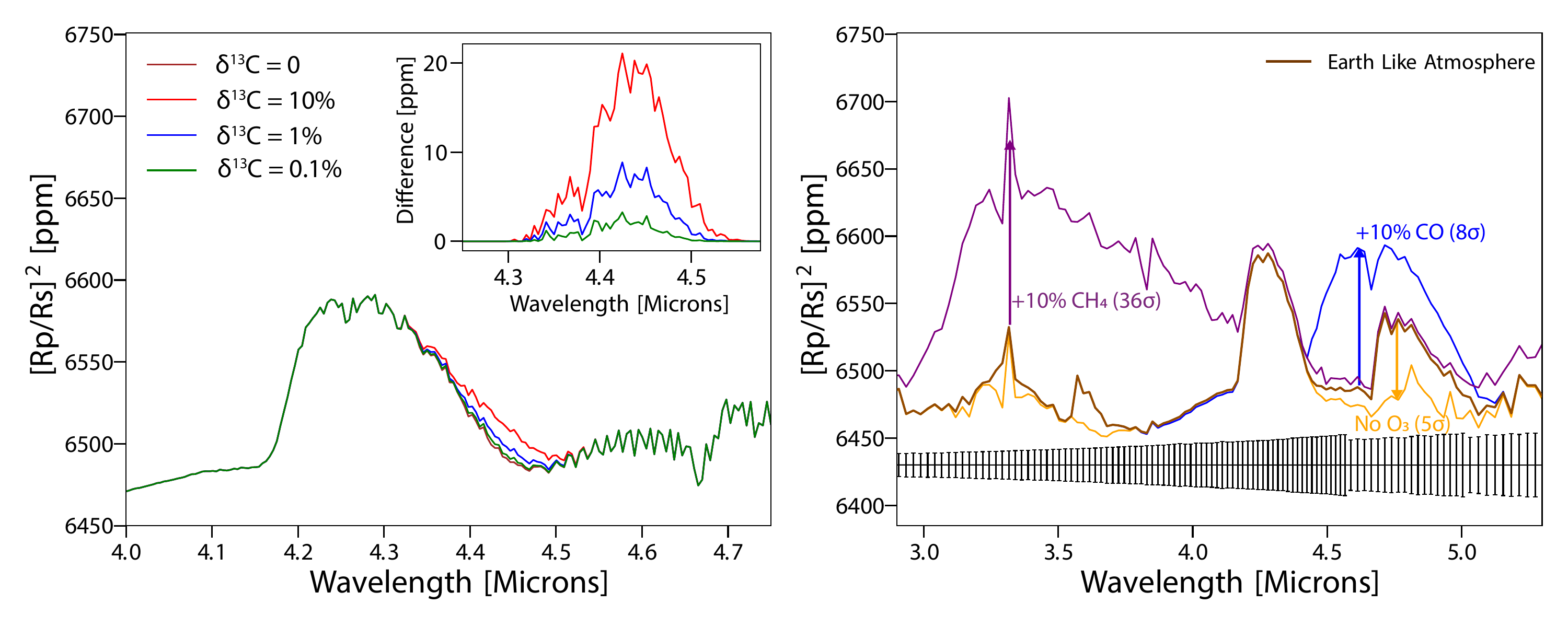}
\caption{Disentangling between the water- or life-based origin of a CO$_2$-depletion (``dCO$_2$'') habiosignature. Left: Effect of different $^{13}$C-to-$^{12}$C ratios ($\delta^{13}$C) on TRAPPIST-1~f transmission spectrum. Effects seen are limited to up to 15ppm, implying that the detecting the isotopic fractionation of carbon in the atmosphere of a temperate terrestrial planet with JWST is out of reach (requires $>$300 transits). Right: Effect of O$_3$ depletion and CH$_4$ and CO 10\% enrichment on TRAPPIST-1~f transmission spectrum. The spectra and their associated 1-$\sigma$ errorbar (black) are shown for a resolving power of R$\sim$130. Detecting ozone in the atmosphere of a temperate terrestrial planet would support the biological origin of a CO$_2$ habiosignature and is within reach with $\sim$ 100 JWST transit observations. 
\label{fig:figure5}}
\end{figure*}

\newpage

\begin{addendum}

 \item[Correspondence] Correspondence and requests for materials
should be addressed to A.H.M.J.T~(email: a.triaud@bham.ac.uk) and J.d.W~(email: jdewit@mit.edu). 

 \item Authors thank M. Gillon and A. Babbin for insightful discussions. B.V.R. thanks the Heising-Simons Foundation for support. AHMJT's research received funding from the European Research Council (ERC) under the European Union's Horizon 2020 research and innovation programme (grant agreement n$^\circ$ 803193/BEBOP).
 
\item[Author Contributions] . 

A.H.M.J.T. and J.d.W. produced the main concepts and led the team behind this Perspective. Every author contributed to the writing of this manuscript. F.K., M.T., O.E.J. and M.P. focused their contribution to the geological discussion of the paper. M.T., J.J.P., A.G., S.S., and F.S. particularly contributed to the atmospheric and biosignature aspects of the paper. B.V.R. and P.N. mostly contributed to the discussion and figures related to observational aspects.

\item[Competing Interests] The authors declare that they have no competing financial interests.

\item[Data Availability]

n/a

\item[Code Availability] 

The JWST photon uncertainties in \autoref{fig:figure4} are calculated using PandExo \cite{batalha2017}.
Systematic uncertainties due to stellar contamination are calculated via a Monte Carlo approach using the temperatures and covering fractions of the two dominant photospheric components inferred from HST observations \cite{Garcia2022}. We allow the covering fraction of the lesser component to vary uniformly from 0\% to double the reported covering fraction, calculate the corresponding stellar contamination signal following ref.~\cite{rackham2018}, and report the standard deviation of the contamination signal for 10,000 such realizations as the systematic uncertainty. We draw the component spectra from the DRIFT-PHOENIX grid \cite{Witte2011} using \texttt{speclib}\footnote{\url{https://github.com/brackham/speclib}} \cite{speclib}. The total uncertainties are the quadrature sums of the photon and systematic uncertainties. 
\texttt{TIERRA}\footnote{\url{https://github.com/disruptiveplanets/tierra}} is a publicly available 1D transmission model written in python. It was originally introduced in ref\cite{Niraula2022} and has been used to create transmission spectra models for Earth-like atmospheres assuming isothermal profiles in \autoref{fig:figure1}, \autoref{fig:figure3} and \autoref{fig:figure5}.

\end{addendum}

\section*{References}





\end{document}